\documentclass[onecollarge,natbib]{svjour2}
\bibpunct{[}{]}{;}{n}{}{,} 
\smartqed  
\usepackage{graphicx}

\usepackage[english]{babel}
\usepackage[utf8x]{inputenc}

\usepackage{latexsym}
\usepackage{amsmath}
\usepackage{amssymb}
\usepackage{amsfonts}
\usepackage{cancel}
\usepackage{bm}
\usepackage{pzccal}

\journalname{Few-Body Systems}

\begin{document}

\title{Elastic and Transition Form Factors in DSEs
}


\author{Jorge Segovia}


\institute{Jorge Segovia \at
           Instituto Universitario de F\'isica Fundamental y Matem\'aticas 
(IUFFyM) \\ 
           Universidad de Salamanca \\
           Plaza de la Merced 1-4, 37008 Salamanca, Spain \\
           Tel.: +34-923-294434 \\
           Fax:  +34-923-294584 \\
           \email{segonza@usal.es} 
}

\date{ Received: / Accepted: }

\maketitle

\begin{abstract}
A symmetry preserving framework for the study of continuum Quantum 
Chromodynamics (QCD) is obtained from a truncated solution of the QCD equations 
of motion or QCD's Dyson-Schwinger equations (DSEs). A nonperturbative solution 
of the DSEs enables the study of, e.g., hadrons as composites of dressed-quarks 
and dressed-gluons, the phenomena of confinement and dynamical chiral symmetry 
breaking (DCSB), and therefrom an articulation of any connection between them. 
It is within this context that we present a unified study of Nucleon, Delta and 
Roper elastic and transition form factors, and compare predictions made using a 
framework built upon a Faddeev equation kernel and interaction vertices that 
possess QCD-like momentum dependence with results obtained using a 
symmetry-preserving treatment of a vector$\,\otimes\,$vector 
contact-interaction.
\keywords{
Dyson-Schwinger equations \and
elastic and transition electromagnetic form factors \and
nucleon resonances
}
\end{abstract}


\vspace*{-0.50cm}
\section{Introduction}
\label{sec:intro}

Nonperturbative QCD poses significant challenges. Primary amongst them is a 
need to chart the behaviour of QCD's running coupling and masses into the 
domain of infrared momenta. Contemporary theory is incapable of solving this 
problem alone but a collaboration with experiment holds a promise for progress. 
This effort can benefit substantially by exposing the structure of nucleon 
excited states and measuring the associated transition form factors at large 
momentum transfers~\cite{Aznauryan:2012ba}. Large momenta are needed in order 
to pierce the meson-cloud that, often to a significant extent, screens the 
dressed-quark core of all baryons~\cite{Roberts:2011rr, Kamano:2013iva}; and it 
is via the $Q^2$ evolution of form factors that one gains access to the running 
of QCD's coupling and masses from the infrared into the 
ultraviolet~\cite{Cloet:2013gva,Chang:2013nia}.

A unified QCD-based description of elastic and transition form factors 
involving the nucleon and its resonances has acquired additional significance 
owing to substantial progress in the extraction of transition electrocouplings, 
$g_{{\rm v}NN^\ast}$, from meson electroproduction data, obtained primarily 
with the CLAS detector at the Thomas Jefferson National Accelerator Facility 
(JLab). The electrocouplings of all low-lying $N^\ast$ states with mass 
less-than $1.6\,{\rm GeV}$ have been determined via independent analyses of 
$\pi^+ n$, $\pi^0p$ and $\pi^+ \pi^- p$ exclusive 
channels~\cite{Agashe:2014kda,Mokeev:2012vsa}; and preliminary results for the 
$g_{{\rm v}NN^\ast}$ electrocouplings of most high-lying $N^\ast$ states with 
masses below $1.8\,{\rm GeV}$ have also been obtained from CLAS meson 
electroproduction data~\cite{Aznauryan:2012ba,Mokeev:2013kka}.

Many new insights have been revealed in a series of recent 
articles~\cite{Segovia:2013rca, Segovia:2013uga, Segovia:2014aza, 
Segovia:2015ufa, Segovia:2015hra} focused on the calculation of the Nucleon, 
Delta and Roper elastic and transition form factors using a widely-used 
leading-order (rainbow-ladder) truncation of QCD's Dyson-Schwinger equations and 
comparing results between a QCD-based framework and a vector$\,\otimes\,$vector 
contact interaction. It is our purpose here reviewing some of the most important 
outcomes and refer to the interested reader to the original works for details.


\begin{figure}[!t]
\begin{center}
\hspace*{0.50cm}
\includegraphics[clip,width=0.40\textwidth,height=0.18\textheight]
{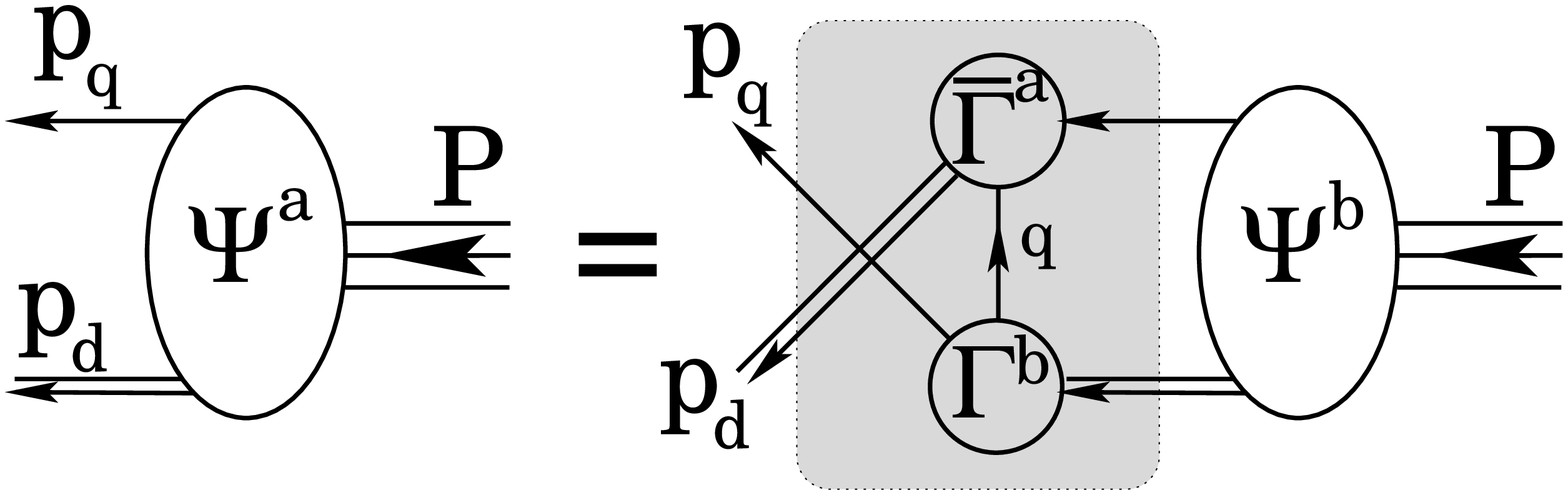} 
\hspace*{1.00cm}
\includegraphics[clip,width=0.40\textwidth,height=0.20\textheight]
{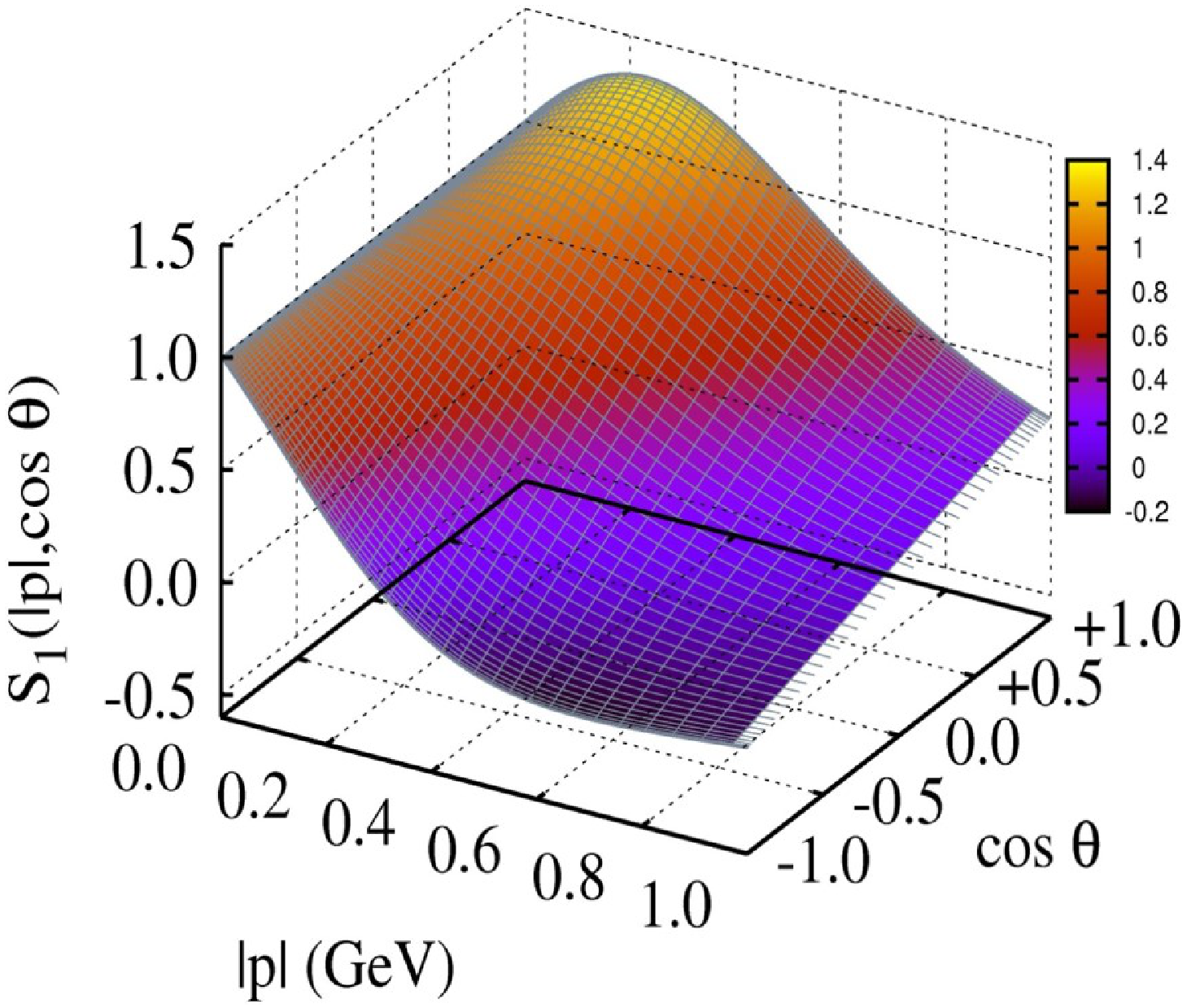}
\caption{\label{fig:Faddeev} {\it Left panel:} Poincar\'e covariant 
Faddeev equation. $\Psi$ is the Faddeev amplitude for a baryon of total 
momentum $P= p_q + p_d$, where $p_{q,d}$ are, respectively, the momenta of the 
quark and diquark within the bound-state. The shaded area demarcates the 
Faddeev equation kernel: {\it single line}, dressed-quark propagator; $\Gamma$, 
diquark correlation amplitude; and {\it double line}, diquark propagator.
{\it Right panel:} Dominant piece in the nucleon's eight-component 
Poincar\'e-covariant Faddeev amplitude: $S_1(|p|,\cos\theta)$. In the nucleon 
rest frame, this term describes that piece of the quark--scalar-diquark 
relative momentum correlation which possesses zero intrinsic quark-diquark 
orbital angular momentum, i.e. $L=0$ before the propagator lines are 
reattached to form the Faddeev wave function. Referring to 
Fig.~\ref{fig:Faddeev}, $p= P/3-p_q$ and $\cos\theta = p\cdot P/\sqrt{p^2 
P^2}$. The amplitude is normalised such that its $U_0$ Chebyshev moment is 
unity at $|p|=0$.
}
\vspace*{-0.50cm}
\end{center}
\end{figure}


\vspace*{-0.50cm}
\section{Baryon structure}
\label{sec:Baryons}

The existence of strong diquark correlations inside baryons is a dynamical 
prediction of Faddeev equation studies based on the observation that the 
attractive nature of quark-antiquark correlations in a colour-singlet meson is 
also attractive for $\bar{3}_{c}$ quark-quark correlations within a 
colour-singlet baryon~\cite{Cahill:1988dx}. 

In the quark$+$diquark picture, baryons are described by the Poincar\'e 
covariant Faddeev equation depicted in the left panel of Fig.~\ref{fig:Faddeev}. 
Two main contributions appear in the binding energy: i) the formation of tight 
diquark correlations and ii) the quark exchange depicted in the shaded area of 
the left panel of Fig.~\ref{fig:Faddeev}\footnote{Whilst an explicit three-body 
term might affect fine details of baryon structure, the dominant effect of 
non-Abelian multi-gluon vertices is expressed in the formation of diquark 
correlations~\cite{Eichmann:2009qa}.}. This exchange ensures that diquark 
correlations within the baryon are fully dynamical: no quark holds a special 
place because each one participates in all diquarks to the fullest extent 
allowed by its quantum numbers. Attending to the quantum numbers of the nucleon 
and Roper, scalar and pseudovector diquark correlations are dominant whereas 
only pseudovector ones are present inside the $\Delta$-baryon.

The quark$+$diquark structure of the nucleon is elucidated in the right panel 
of Fig.~\ref{fig:Faddeev}, which depicts the leading component of its Faddeev 
amplitude: with the notation of Ref.~\cite{Segovia:2014aza}, 
$S_1(|p|,\cos\theta)$, computed using the Faddeev kernel described therein. 
This function describes a piece of the quark$+$scalar-diquark relative momentum 
correlation. Notably, in this solution of a realistic Faddeev equation there is 
strong variation with respect to both arguments. Support is concentrated in 
the forward direction, $\cos\theta >0$, so that alignment of $p$ and $P$ is 
favoured; and the amplitude peaks at $(|p|\simeq M_N/6,\cos\theta=1)$, whereat 
$p_q \sim p_d \sim P/2$ and hence the natural relative momentum is zero. In the 
antiparallel direction, $\cos\theta<0$, support is concentrated at $|p|=0$, 
i.e. $p_q \sim P/3$, $p_d \sim 2P/3$.

\begin{figure}[!t]
\centerline{
\includegraphics[clip,width=0.40\textwidth]{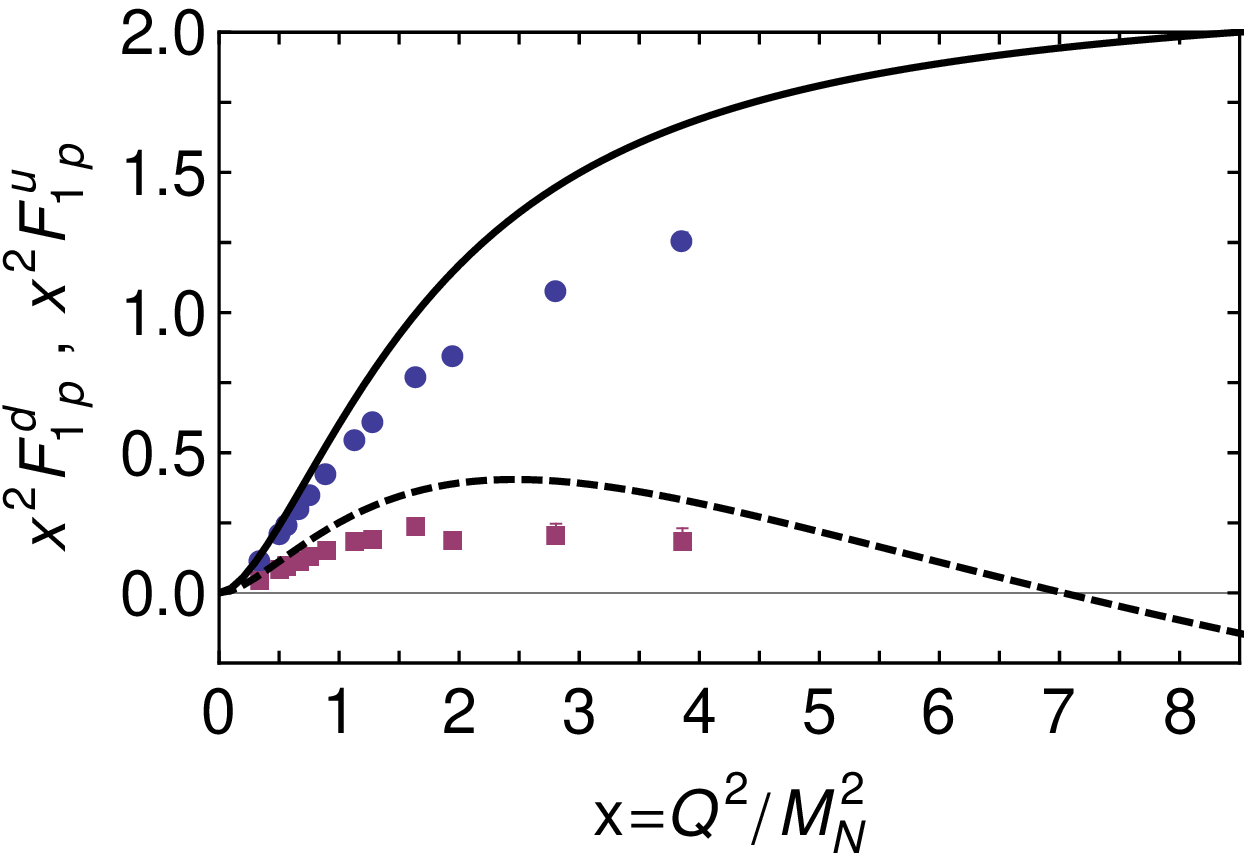}
\hspace*{1.00cm}
\includegraphics[clip,width=0.40\textwidth]{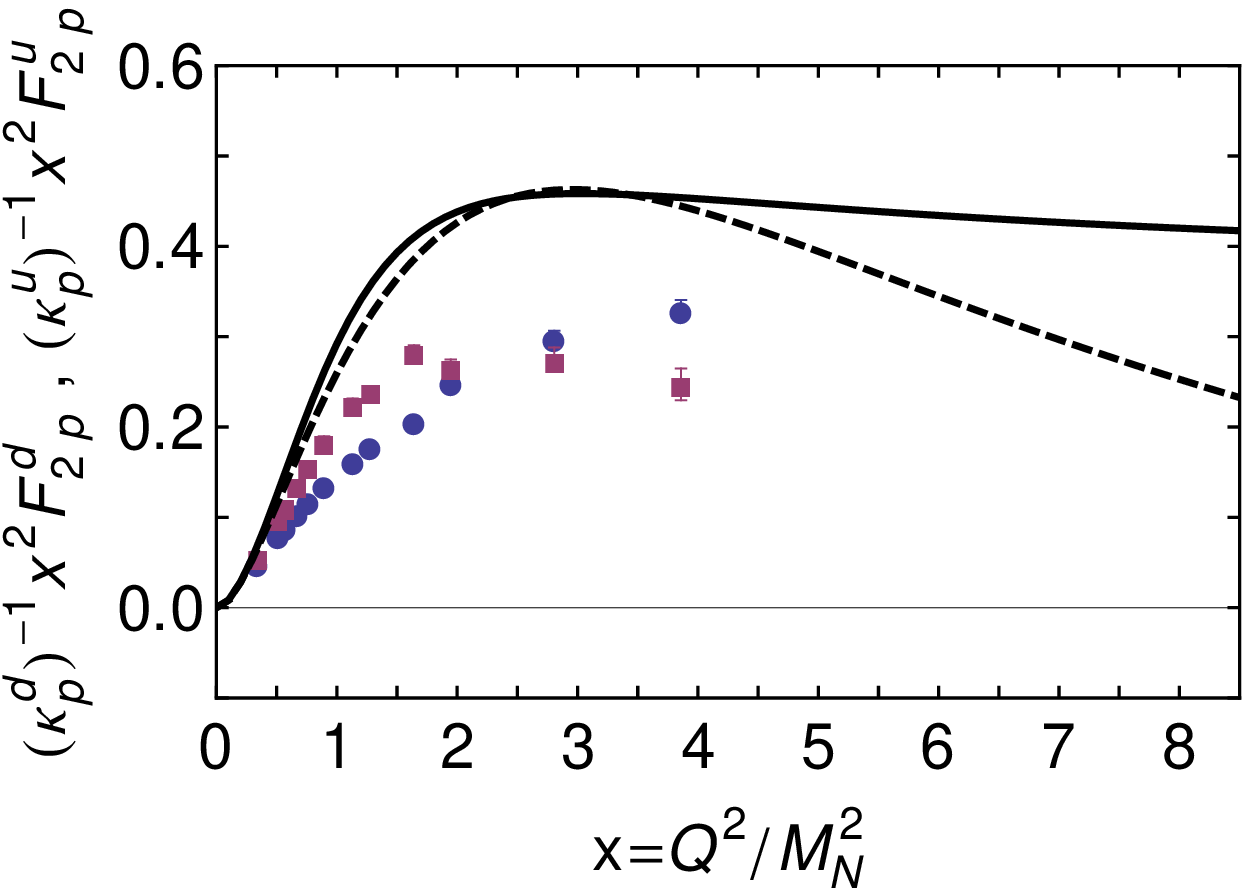}}
\caption{\label{fig:F1F2fla1} {\it Left panel:} Flavour separation of the 
proton's Dirac form factor as a function of $x=Q^2/M_N^2$. The results have 
been obtained using a framework built upon a Faddeev equation kernel and 
interaction vertices that possess QCD-like momentum dependence. The solid-curve 
is the $u$-quark contribution, and the dashed-curve is the $d$-quark 
contribution. Experimental data taken from Ref.~\protect\cite{Cates:2011pz} and 
references therein: circles -- $u$-quark; and squares -- $d$-quark. {\it Right 
panel:} Same for Pauli form factor.
%
}
\vspace*{-0.25cm}
\end{figure}

The strong diquark correlations must be evident in many physical observables. 
We focus our attention on the flavour separated versions of the Dirac a Pauli 
form factors of the nucleon. Figure~\ref{fig:F1F2fla1} displays the proton's 
flavour separated Dirac and Pauli form factors. The salient features of the 
data are: the $d$-quark contribution to $F_1^p$ is far smaller than the 
$u$-quark contribution; $F_2^d/\kappa_d>F_2^u/\kappa_u$ on $x<2$ but this 
ordering is reversed on $x>2$; and in both cases the $d$-quark contribution 
falls dramatically on $x>3$ whereas the $u$-quark contribution remains roughly 
constant. Our calculations are in semi-quantitative agreement with the 
empirical data.

It is natural to seek an explanation for the pattern of behaviour in 
Fig.~\ref{fig:F1F2fla1}. We have mentioned that the proton contains scalar and 
pseudovector diquark correlations. The dominant piece of its Faddeev wave 
function is $u[ud]$; namely, a $u$-quark in tandem with a $[ud]$ scalar 
correlation, which produces $62\%$ of the proton's normalisation. If this were 
the sole component, then photon--$d$-quark interactions within the proton would 
receive a $1/x$ suppression on $x>1$, because the $d$-quark is sequestered in a 
soft correlation, whereas a spectator $u$-quark is always available to 
participate in a hard interaction. At large $x=Q^2/M_N^2$, therefore, scalar 
diquark dominance leads one to expect $F^d \sim F^u/x$.  Available data are 
consistent with this prediction but measurements at $x>4$ are necessary for 
confirmation.

\begin{figure}[!t]
\begin{center}
\begin{tabular}{ll}
\includegraphics[clip,height=0.20\textheight,width=0.43\textwidth]
{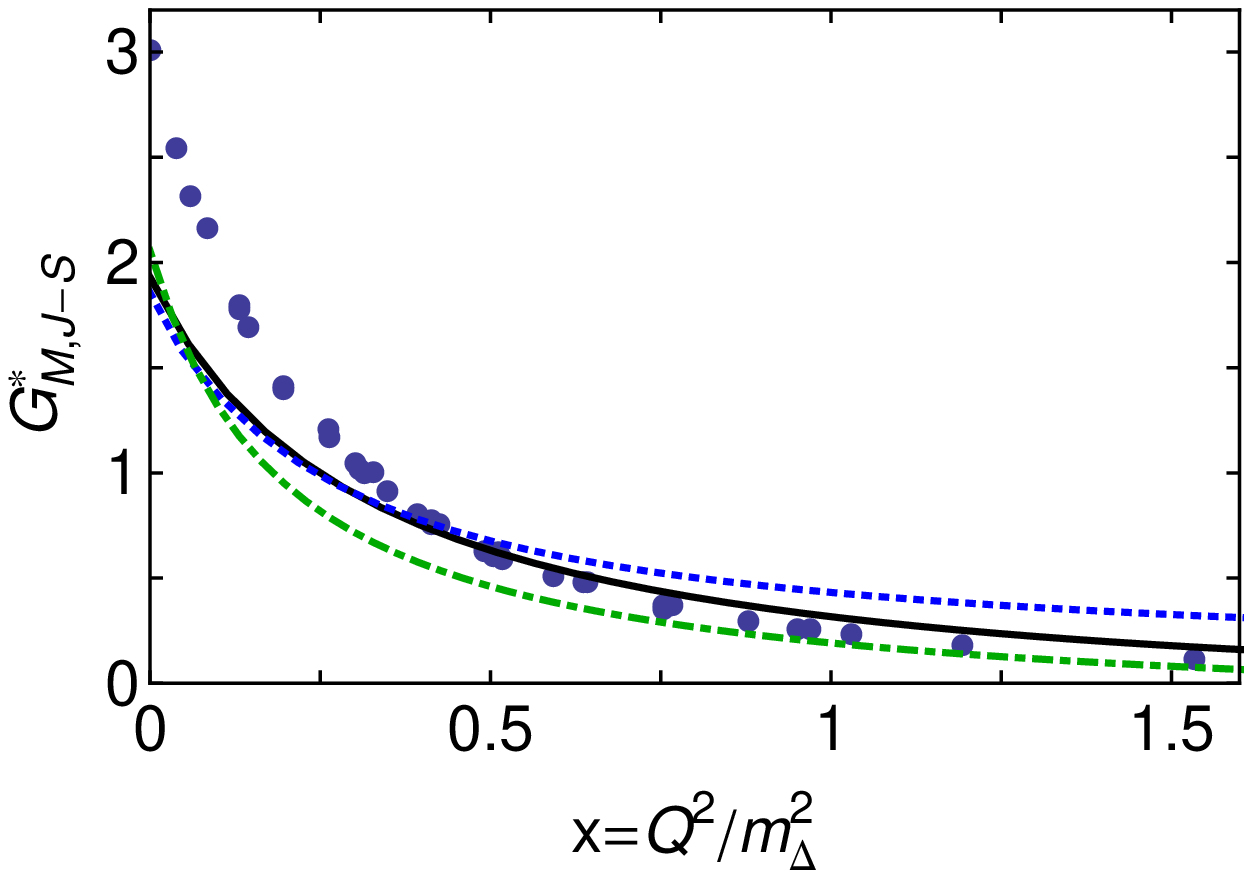}
& 
\includegraphics[clip,height=0.2025\textheight,width=0.45\textwidth]
{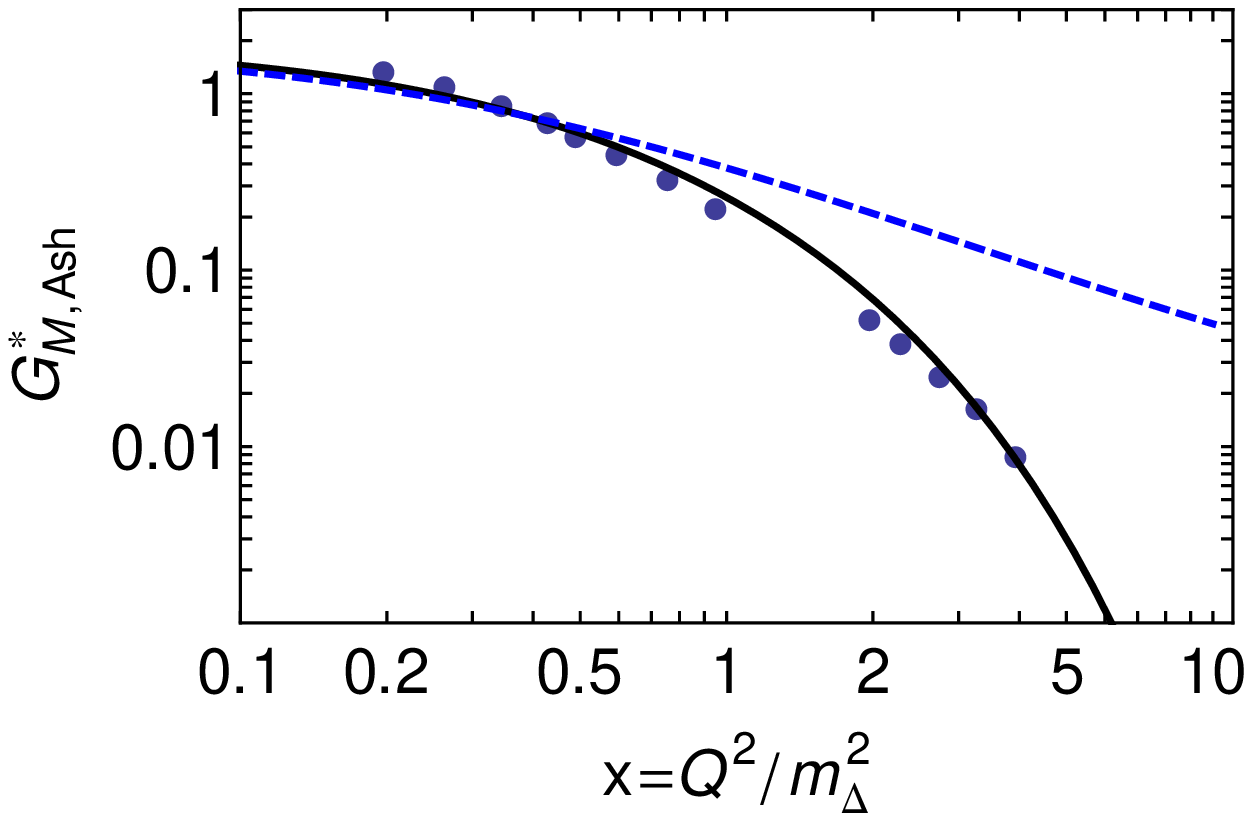} \\
\hspace*{-0.50cm}
\includegraphics[clip,height=0.20\textheight,width=0.455\textwidth]
{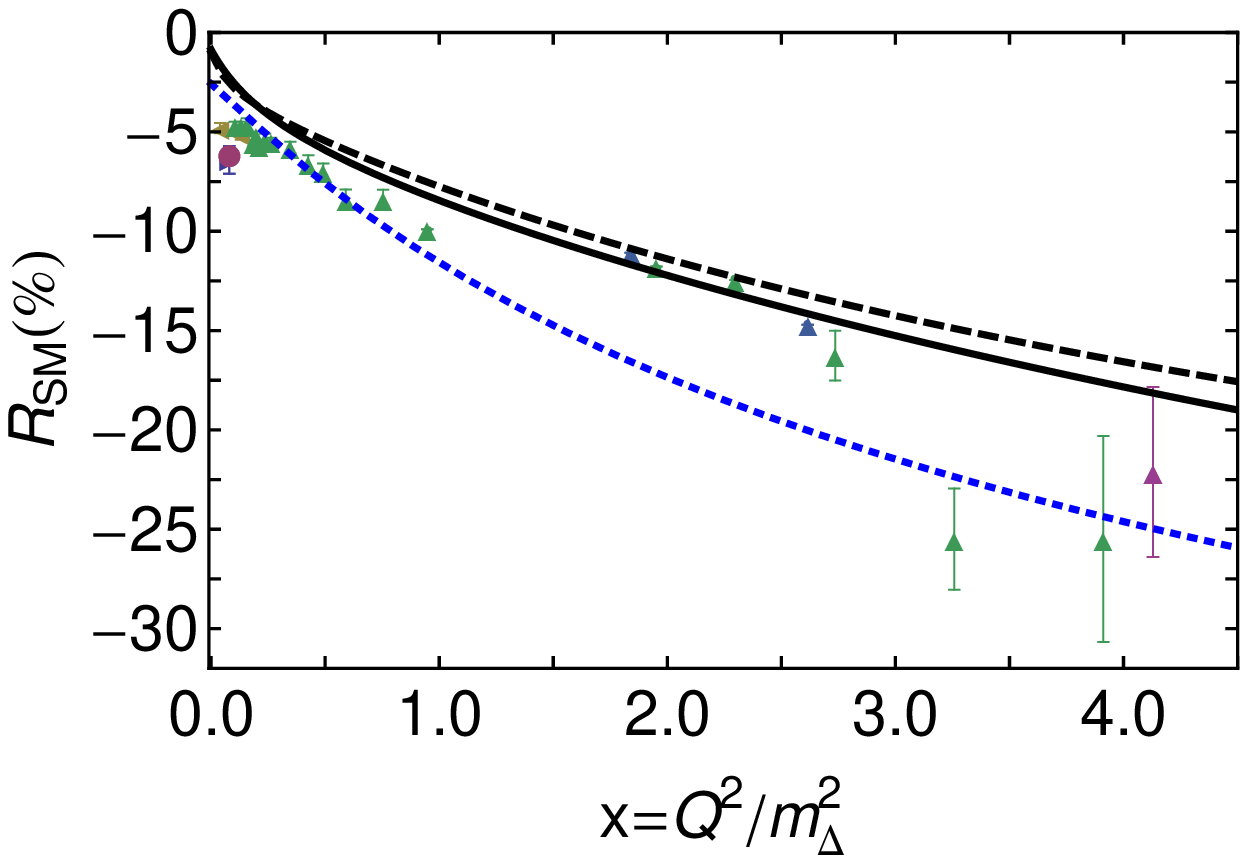}
& 
\includegraphics[clip,height=0.20\textheight,width=0.45\textwidth]
{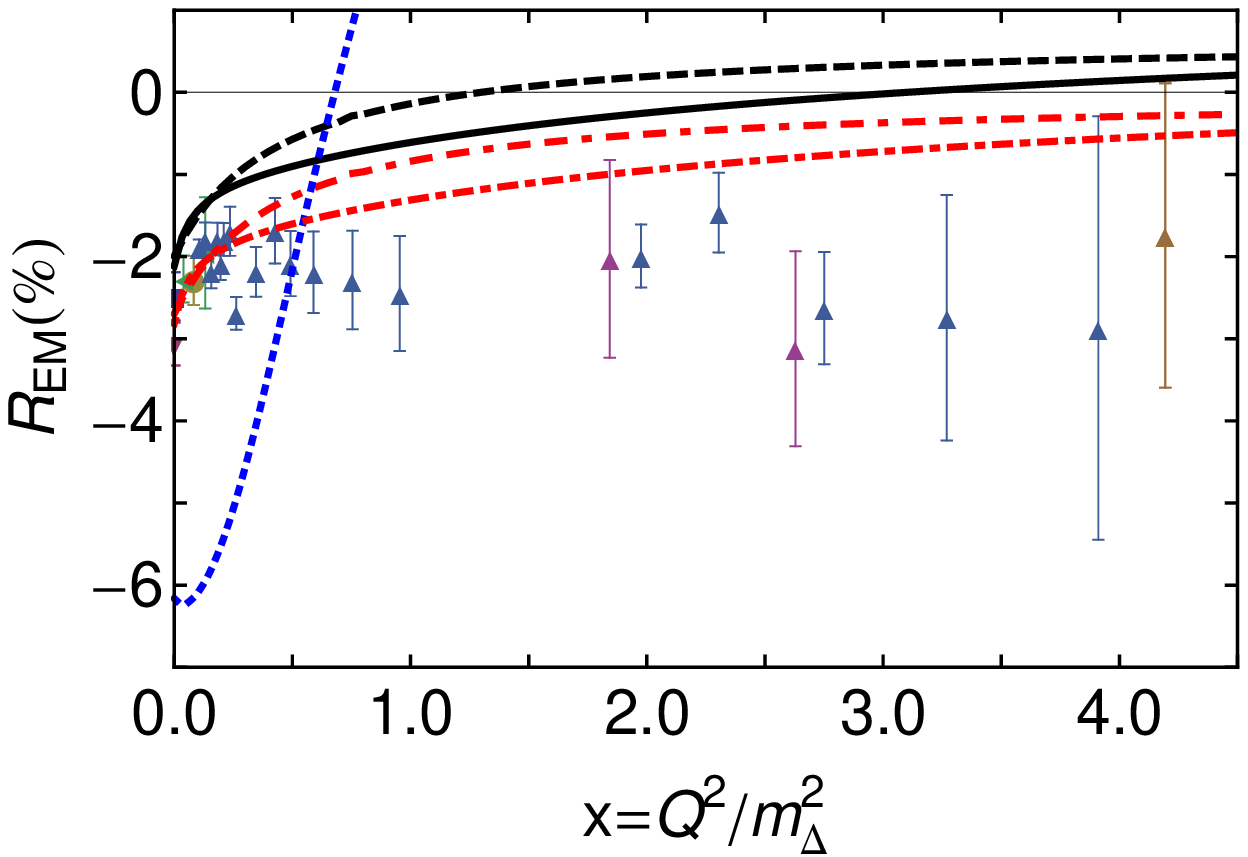}
\end{tabular}
\caption{\label{fig:NucDel} 
\emph{Upper-left panel} -- $G_{M,J-S}^{\ast}$ result obtained with QCD-based 
interaction (solid, black) and with contact-interaction (CI) (dotted, blue); 
The green dot-dashed curve is the dressed-quark core contribution inferred 
using SL-model~\protect\cite{JuliaDiaz:2006xt}.
\emph{Upper-right panel} -- $G_{M,Ash}^{\ast}$ result obtained with QCD-based 
interaction (solid, black) and with CI (dotted, blue).
\emph{Lower-left panel} -- $R_{SM}$ prediction of QCD-based kernel including 
dressed-quark anomalous magnetic moment (DqAMM) (black, solid), nonincluding 
DqAMM (black, dashed), and CI result (dotted, blue).
\emph{Lower-right panel} -- $R_{EM}$ prediction obtained with QCD-kindred 
framework (solid, black); same input but without DqAMM (dashed, black); these 
results renormalised (by a factor of $1.34$) to agree with experiment at $x=0$ 
(dot-dashed, red - zero at $x\approx 14$; and dot-dash-dashed, red, zero 
at $x\approx 6$); and CI result (dotted, blue).
The data in the panels are from references that can be found 
in~\protect\cite{Segovia:2014aza}.
}
\vspace*{-0.50cm}
\end{center}
\end{figure}


\vspace*{-0.50cm}
\section{The \mbox{\boldmath $\gamma^\ast N \to \Delta$} Transition}
\label{sec:FFnucdel}

The electromagnetic $\gamma^{\ast}N\to \Delta$ transition is described by 
three Poincar\'e-invariant form factors~\cite{Jones:1972ky}: magnetic-dipole, 
$G_{M}^{\ast}$, electric quadrupole, $G_{E}^{\ast}$, and Coulomb (longitudinal) 
quadrupole, $G_{C}^{\ast}$; that can be extracted in the Dyson-Schwinger 
approach by a sensible set of projection operators~\cite{Eichmann:2011aa}. The 
following ratios
\begin{equation}
R_{\rm EM} = -\frac{G_E^{\ast}}{G_M^{\ast}}, \qquad
R_{\rm SM} = - \frac{|\vec{Q}|}{2 m_\Delta} \frac{G_C^{\ast}}{G_M^{\ast}}\,,
\label{eq:REMSM}
\end{equation}
are often considered because they can be read as measures of the deformation of 
the hadrons involved in the reaction and how such deformation influences the 
structure of the transition current.

The upper-left panel of Fig.~\ref{fig:NucDel} displays the magnetic transition 
form factor in the Jones-Scadron convention. Our prediction obtained with a 
QCD-based kernel agrees with the data on $x\gtrsim 0.4$, and a similar 
conclusion can be inferred from the contact interaction result. On the other 
hand, both curves disagree markedly with the data at infrared momenta. This is 
explained by the similarity between these predictions and the bare result 
determined using the Sato-Lee (SL) dynamical meson-exchange 
model~\cite{JuliaDiaz:2006xt}. The SL result supports a view that the 
discrepancy owes to omission of meson-cloud effects in the DSEs' computations. 
An exploratory study of the effect of pion-cloud contributions to the mass of 
the nucleon and the $\Delta$-baryon has been performed within a DSEs' framework 
in Ref.~\cite{Sanchis-Alepuz:2014wea}.

Presentations of the experimental data associated with the magnetic transition 
form factor typically use the Ash convention. This comparison is depicted in 
the upper-right panel of Fig.~\ref{fig:NucDel}. One can see that the difference 
between form factors obtained with the QCD-kindred and CI frameworks increases 
with the transfer momentum. Moreover, the normalized QCD-kindred curve is in 
fair agreement with the data, indicating that the Ash form factor falls 
unexpectedly rapidly mainly for two reasons. First: meson-cloud effects provide 
up-to $35\%$ of the form factor for $x \lesssim 2$; these contributions are very 
soft; and hence they disappear quickly. Second: the additional kinematic factor 
$\sim 1/\sqrt{Q^2}$ that appears between Ash and Jones-Scadron conventions and 
provides material damping for $x\gtrsim 2$.

Our predictions for the ratios in Eq.~(\ref{eq:REMSM}) are depicted in the 
lower panels of Fig.~\ref{fig:NucDel}. The left panel displays the Coulomb 
quadrupole ratio. Both the prediction obtained with QCD-like propagators and 
vertices and the contact-interaction result are broadly consistent with 
available data. This shows that even a contact-interaction can produce 
correlations between dressed-quarks within Faddeev wave-functions and related
features in the current that are comparable in size with those observed 
empirically. Moreover, suppressing the dressed-quark anomalous magnetic moment 
(DqAMM) in the transition current has little impact. These remarks highlight 
that $R_{SM}$ is not particularly sensitive to details of the Faddeev kernel 
and transition current.

This is certainly not the case with $R_{\rm EM}$. The differences between the 
curves displayed in the lower-right panel in Fig.~\ref{fig:NucDel} show that 
this ratio is a particularly sensitive measure of diquark and orbital angular 
momentum correlations. The contact-interaction result is inconsistent with 
data, possessing a zero that appears at a rather small value of $x$. On the 
other hand, predictions obtained with QCD-like propagators and vertices can be 
viable. We have presented four variants, which differ primarily in the location 
of the zero that is a feature of this ratio in all cases we have considered. 
The inclusion of a DqAMM shifts the zero to a larger value of $x$. Given the 
uniformly small value of this ratio and its sensitivity to the DqAMM, we judge 
that meson-cloud affects must play a large role on the entire domain that is 
currently accessible to experiment.


\begin{figure}[!t]
\begin{center}
\includegraphics[clip,height=0.18\textheight,width=0.40\textwidth]
{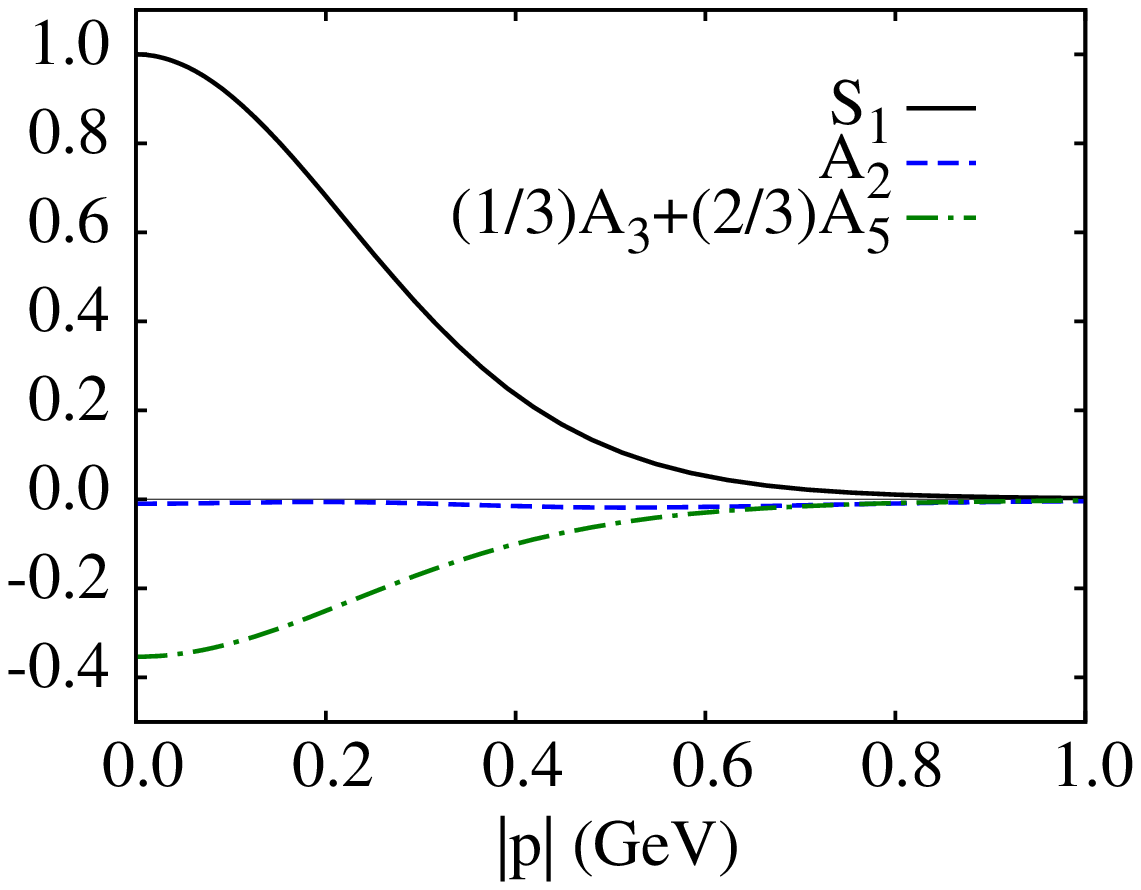}
\hspace*{1.00cm}
\includegraphics[clip,height=0.18\textheight,width=0.40\textwidth]
{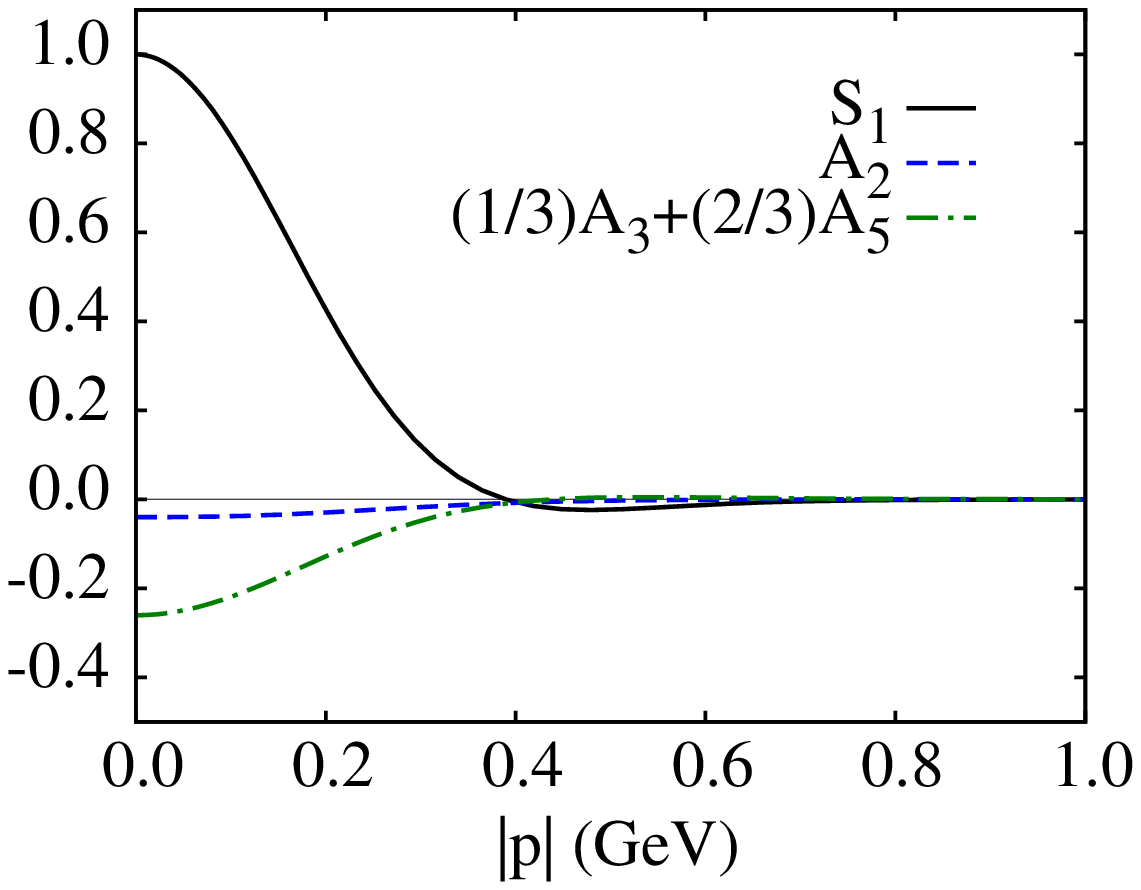}
\caption{\label{fig:NucRop_v1} \emph{Left panel}. Zeroth Chebyshev moment of 
all $S$-wave components in the nucleon's Faddeev wave function. \emph{Right 
panel}. Kindred functions for the first excited state. Legend: $S_1$ is 
associated with the baryon's scalar diquark; the other two curves are 
associated 
with the axial-vector diquark; and the normalisation is chosen such that 
$S_1(0)=1$.}
\vspace*{-0.50cm}
\end{center}
\end{figure}

\begin{figure}[!t]
\begin{minipage}[t]{\textwidth}
\begin{minipage}{0.49\textwidth}
\centerline{\includegraphics[clip,width=0.85\linewidth]{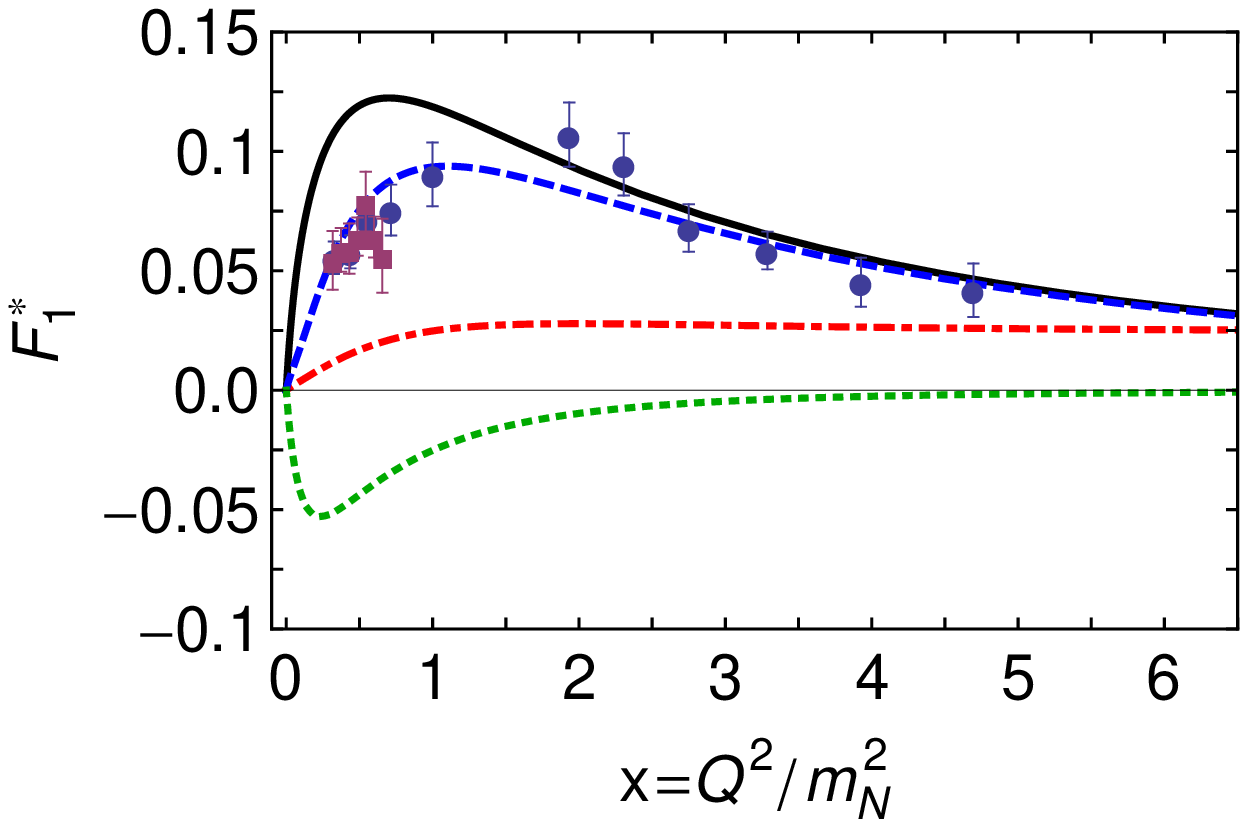}}
\end{minipage}
\begin{minipage}{0.49\textwidth}
\centerline{\includegraphics[clip,width=0.85\linewidth]{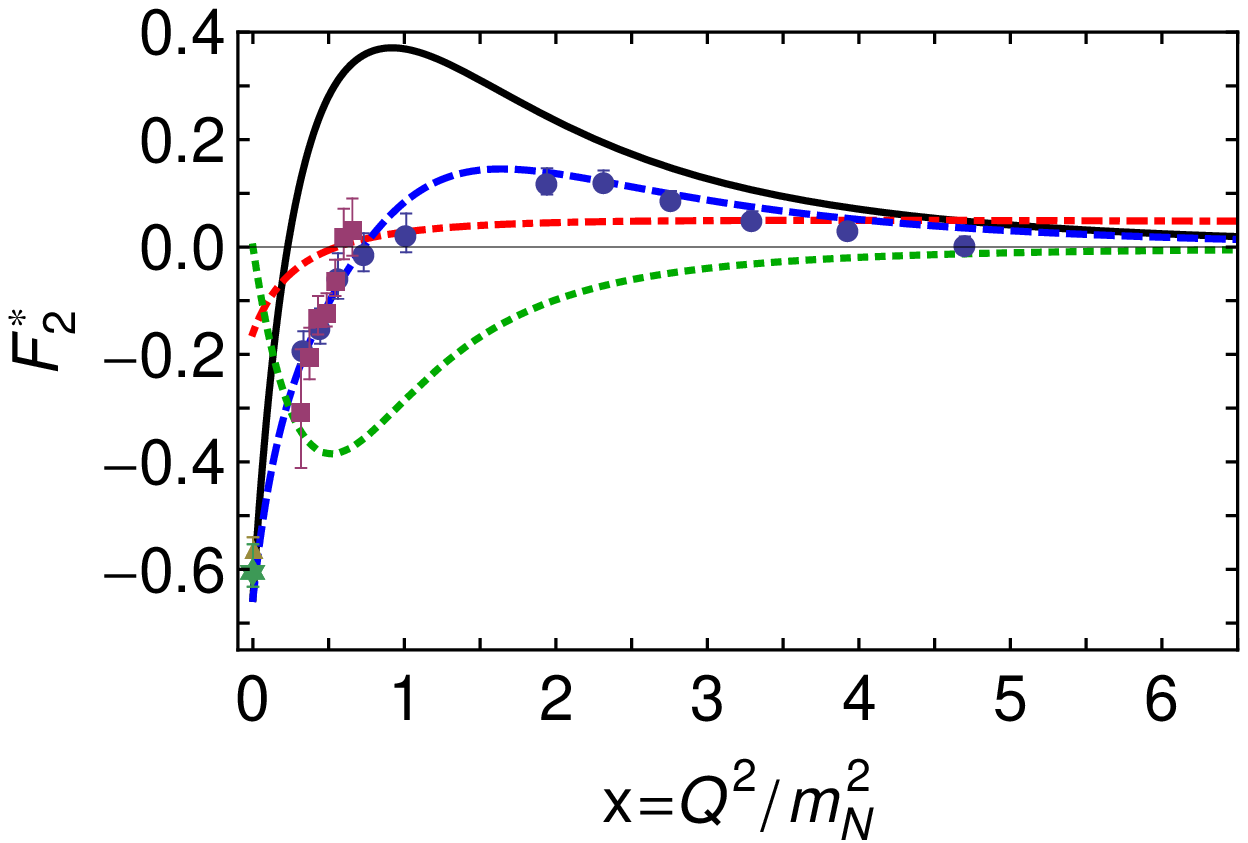}}
\end{minipage}
\end{minipage}
\caption{\label{fig:NucRop_v2} \emph{Left} -- Dirac transition form factor, 
$F_{1}^{\ast}(x)$, $x=Q^2/m_N^2$. Solid (black) curve, QCD-kindred prediction; 
dot-dashed (red) curve, contact-interaction result; dotted (green) curve, 
inferred meson-cloud contribution; and dashed (blue) curve, anticipated 
complete result. \emph{Right} -- Pauli transition form factor, 
$F_{2}^{\ast}(x)$, with same legend. Data in both panels: circles 
(blue)~\cite{Aznauryan:2009mx}; triangle (gold)~\cite{Dugger:2009pn}; squares 
(purple)~\cite{Mokeev:2012vsa}; and star (green)~\cite{Agashe:2014kda}.}
\vspace*{-0.50cm}
\end{figure}


\vspace*{-0.50cm}
\section{The \mbox{\boldmath $\gamma^\ast N \to Roper$} Transition}
\label{sec:Roper}

Jefferson Lab experiments~\cite{Aznauryan:2011qj, Aznauryan:2009mx, 
Dugger:2009pn, Mokeev:2012vsa} have yielded precise nucleon-Roper ($N\to R$) 
transition form factors and thereby exposed the first zero seen in any hadron 
form factor or transition amplitude. It has also attracted much theoretical 
attention; but Ref.~\cite{Segovia:2015hra} provides the first continuum 
treatment of this problem using the power of relativistic quantum field theory. 
That study begins with a computation of the mass and wave function of the 
proton and its first radial excitation. The masses are (in GeV): $M_{\rm 
nucleon\,(N)} = 1.18$ and $M_{\rm nucleon-excited\,(R)}=1.73$.
These values correspond to the locations of the two lowest-magnitude $J^P=1/2^+$ 
poles in the three-quark scattering problem. The associated residues are the 
Faddeev wave functions, which depend upon $(p^2,p\cdot P)$, where $p$ is the 
quark-diquark relative momentum. Fig.~\ref{fig:NucRop_v1} depicts the zeroth 
Chebyshev moment of all $S$-wave components in that wave function. The 
appearance of a single zero in $S$-wave components of the Faddeev wave function 
associated with the first excited state in the three dressed-quark scattering 
problem indicates that this state is a radial excitation.

The empirical values of the pole locations for the first two states in the 
nucleon channel are~\cite{Suzuki:2009nj}: $0.939\,{\rm GeV}$ and $1.36 - i 
\, 0.091\,{\rm GeV}$, respectively. At first glance, these values appear 
unrelated to those obtained within the DSEs framework.
However, deeper consideration reveals~\cite{Eichmann:2008ae, Eichmann:2008ef} 
that the kernel in the Faddeev equation omits all those resonant contributions 
which may be associated with the meson-baryon final-state interactions that are 
resummed in dynamical coupled channels models in order to transform a 
bare-baryon into the observed state~\cite{Suzuki:2009nj, Kamano:2013iva}. This 
Faddeev equation should therefore be understood as producing the dressed-quark 
core of the bound-state, not the completely-dressed and hence observable 
object. Crucial, therefore, is a comparison between the quark-core mass and the 
value determined for the mass of the meson-undressed bare-Roper in 
Ref.~\cite{Suzuki:2009nj} which is $1.76\,{\rm GeV}$.

The transition form factors are displayed in Fig.~\ref{fig:NucRop_v2}. The 
results obtained using QCD-derived propagators and vertices agree with the data 
on $x\gtrsim 2$. The contact-interaction result simply disagree both 
quantitatively and qualitatively with the data. Therefore, experiment is 
evidently a sensitive tool with which to chart the nature of the quark-quark 
interaction and hence discriminate between competing theoretical hypotheses.

The mismatch between the DSE predictions and data on $x\lesssim 2$ is due to 
Meson-cloud contributions that are expected to be important on this domain. An 
inferred form of that contribution is provided by the dotted (green) curves in 
Fig.~\ref{fig:NucRop_v2}. These curves have fallen to just 20\% of their 
maximum value by $x=2$ and vanish rapidly thereafter so that the DSE 
predictions alone remain as the explanation of the data. Importantly, the 
existence of a zero in $F_{2}^{\ast}$ is not influenced by meson-cloud effects, 
although its precise location is.


\vspace*{-0.50cm}
\section{Conclusions}
\label{sec:conclusions}
\vspace*{-0.20cm}

We described a unified study of Nucleon, Delta and Roper elastic and transition 
form factors that compares predictions made by a QCD-kindred framework with 
results obtained using a symmetry-preserving treatment of a 
vector$\,\otimes\,$vector contact-interaction. The comparison emphasises 
that experiment is sensitive to the momentum dependence of the running 
coupling and masses in QCD. Amongst our results, the following are of 
particular interest:
the presence of strong diquark correlations within the nucleon is sufficient to 
understand empirical extractions of the flavour-separated form factors.
In connection with the $\gamma^{\ast}N\to \Delta$ transition, the 
momentum-dependence of the magnetic transition form factor, $G_M^\ast$, matches 
that of $G_M^n$ once the momentum transfer is high enough to pierce the 
meson-cloud; and the electric quadrupole ratio is a keen measure of diquark and 
orbital angular momentum correlations, the zero in which is obscured by 
meson-cloud effects on the domain currently accessible to experiment.
Finally, the Roper resonance is at heart of the nucleon's first radial 
excitation, consisting of a dressed-quark core augmented by a meson cloud that 
reduces its mass by approximately $20\%$. Our analysis shows that a meson-cloud 
obscures the dressed-quark core from long-wavelength probes, but that it is 
revealed to probes with $Q^2 \gtrsim 3 m_N^2$.


\vspace*{-0.30cm}
\begin{acknowledgements}
The material described in this contribution is drawn from work completed in 
collaboration with numerous excellent people, to all of whom I am greatly 
indebted.
I would also like to thank V. Mokeev, R.~Gothe, T.-S.\,H.~Lee and G. Eichmann 
for insightful comments;
and to express my gratitude to the organisers of the {\it Light Cone 2015}, 
whose support helped my participation.
I acknowledges financial support from a postdoctoral IUFFyM contract at 
Universidad de Salamanca, Spain.
\end{acknowledgements}

\vspace*{-0.80cm}



\end{document}